# TIME VARIATIONS OF THE SOLAR NEUTRINO FLUX DATA FROM SAGE AND GALLEX-GNO DETECTORS OBTAINED BY RAYLEIGH POWER SPECTRUM ANALYSIS

P. Raychaudhuri[a] and K. Ghosh[b]

(a)Department of Applied Mathematics, University of Calcutta, 92, A.P.C. Road, Calcutta-700009, INDIA
(b)Department of Mathematics, Dr. B. C. Roy Engineering College, Durgapur- 713206, INDIA
Presenter: K.Ghosh (koushikg123@yahoo.co.uk), ind-ghosh-koushik-abs2-he22-poster

We have used Rayleigh power spectrum analysis of the monthly solar neutrino flux data from (1) SAGE detector during the period from 1st January 1990 to 31st December 2000; (2) SAGE detector during the period from April 1998 to December 2001; (3) GALLEX detector during the period from May 1991 to January 1997; (4) GNO detector during the period from May 1998 to December 2001; (5) GALLEX-GNO detector (combined data) from May 1991 to December 2001 and (6) average of the data from GNO and SAGE detectors during the period from May 1998 to December 2001. (1) exhibits periodicity around 1.3, 4.3, 5.5, 6.3, 7.9, 8.7, 15.9, 18.7, 23.9, 32.9 and 48.7 months. (2) shows periodicity around 1.5, 2.9, 4.5, 10.1 months. For (3) we observe periodicity around 1.7, 18.7 and 26.9 months. For (4) periodicity is seen around 3.5, 5.5, 7.7 and 10.5 months. (5) gives periodicity around 1.7, 18.5, 28.5 and 42.1 months while (6) shows periodicity around 4.3, 6.9, 10.3 and 18.1 months. We have found almost similar periods in the solar flares, sunspot data, solar proton data ($\bar{\epsilon} > 10$ Mev) which indicates that the solar activity cycle may be due to the variable character of nuclear energy generation inside the sun.

## 1. Introduction

Solar neutrino flux detection is very important not only to understand the stellar evolution but also to understand the origin of the solar activity cycle. Recent solar neutrino flux observed by Super-Kamiokande [1] and SNO detectors [2] suggest that solar neutrino flux from $^8$B neutrino and $^3$He+p neutrino from Standard Solar Model (S.S.M.)[3] is at best compatible with S.S.M. calculation if we consider the neutrino oscillation of M.S.W.[4] or if the neutrino flux from the sun is a mixture of two kinds of neutrino i.e. $\nu_e$ and $\nu_\mu$ [5] . Standard Solar Model (S.S.M.) are known to yield the stellar structure to a very good degree of precision but the S.S.M. cannot explain the solar activity cycle, the reason being that this S.S.M. does not include temperature and magnetic variability of the solar core [6,7]. The temperature variability implied a variation of the energy source and from that source of energy magnetic field can be generated which also imply a magnetic variability [7]. The temperature variation is important for the time variation of the solar neutrino flux. So we need a perturbed solar model and it is outlined by Raychaudhuri since 1971[6,7], which may satisfy all the requirements of solar activity cycle with S.S.M. For the support of perturbed solar model we have demonstrated that solar neutrino flux data are fractal in nature [8,9,10]. The excess nuclear energy from the perturbed nature of the solar model transforms into magnetic energy, gravitational energy and thermal energy etc. below the tachocline. The variable nature of magnetic energy induces dynamic action for the generation of solar magnetic field.

Solar neutrino flux data from Homestake [11] detector varies with the solar activity cycle but at present it appears that there is no significant anti correlation of solar neutrino flux data with the sunspot numbers. Many authors analysed the solar neutrino flux data from Homestake detector and have found short-time periodicities around 5, 10, 15, 20, 25 months etc. Raychaudhuri analysed the solar neutrino flux data from SAGE, GALLEX, Superkamiokande and have found that solar neutrino flux data varies with the solar activity cycle and have found periodicities around 5 and 10 months.

The purpose of the present paper is to see whether the SAGE, GALLEX-GNO solar neutrino flux data are variable in nature or not. The observation of a variable nature of solar neutrino would provide significance to our understanding of solar internal dynamics and probably to the requirement of the modification of the Standard Solar Model i.e. a perturbed solar model. In this paper we shall study the monthly solar neutrino flux data from (1) SAGE detector during the period from 1st January 1990 to 31st December 2000 [12]; (2) SAGE detector during the period from April 1998 to December 2001 [12]; (3) GALLEX detector during the period from May 1991 to January 1997 [13]; (4) GNO detector during the period from May 1998 to December 2001 [13]; (5) GALLEX-GNO detector (combined data) from May 1991 to December 2001 [13] and (6) average of the data from GNO and SAGE detector during the period from May 1998 to December 2001 [12, 13]. The obtained data are used for the analysis of periodicity following the method of Rayleigh Power Spectrum Analysis.

## 2. Rayleigh Power Spectrum Analysis

Suppose we want to determine whether n events with angular values of $\{\theta_1, \theta_2, \theta_3, \ldots, \theta_n\}$ are uniformly distributed in angle. We can represent each event as a unit vector

$$\vec{u_i} = \cos\theta_i \cdot \hat{e}_x + \sin\theta_i \cdot \hat{e}_y$$

where $\hat{e}_x$ and $\hat{e}_y$ are unit vectors parallel to the x-axis and the y-axis respectively. The vector sum of these unit vectors is given by [14]

$$\vec{U} = \sum_{i=1}^{n} \cos\theta_i \cdot \hat{e}_x + \sum_{i=1}^{n} \sin\theta_i \cdot \hat{e}_y \qquad (1)$$

The magnitude of this vector divided by the number of events [14]

$$R = (1/n) \left[ \left(\sum_{i=1}^{n} \cos\theta_i\right)^2 + \left(\sum_{i=1}^{n} \sin\theta_i\right)^2 \right]^{1/2} \qquad (2)$$

indicates the uniformity of the distribution. If the events are uniformly distributed R is very close to zero. If on the other hand the events are concentrated around a certain angle, R is close to unity. The direction angle of the vector $\vec{U}$ shows the angle around which the events are concentrated. Bai and Cliver [14] defined the quantity Z as

$$Z = nR^2 = (1/n) \left[ \left(\sum_{i=1}^{n} \cos\theta_i\right)^2 + \left(\sum_{i=1}^{n} \sin\theta_i\right)^2 \right] \qquad (3)$$

for randomly distributed events and the distribution of Z follows[5] $P(Z>K)=\exp(-K)$. They obtained the "RAYLEIGH POWER SPECTRUM" $Z(\nu)$ by setting $\theta_i = 2\pi t_i/T = 2\pi\nu_i$, where $\{t_i\}$ is a set of event occurrence times and T is a variable period.

It is to be noted that Bai and Cliver [14] did not consider the observed data of occurrence. They just considered the set of time of occurrence. Here we have modified the idea of Bai and Cliver [14] where we have considered the observed data as well as the set of time of occurrence. Here we have modified each event as a vector of modulus $|x(t_i)|$ instead of a unit vector considered by Bai and Cliver [14] as

$$\vec{u_i} = x(t_i).\cos\theta_i.\hat{e_x} + x(t_i).\sin\theta_i.\hat{e_y}$$ and the vector sum of these vectors is given by

$$\vec{U} = \sum_{i=1}^{n} x(t_i).\cos\theta_i.\hat{e_x} + \sum_{i=1}^{n} x(t_i)\sin\theta_i.\hat{e_y} \quad (4)$$

Again the magnitude of this vector divided by the number of events is given by

$$R = (1/n)[(\sum_{i=1}^{n} x(t_i)\cos\theta_i)^2 + (\sum_{i=1}^{n} x(t_i)\sin\theta_i)^2]^{1/2} \quad (5)$$

Here the quantity Z is defined as

$$Z = nR^2 = (1/n)[(\sum_{i=1}^{n} x(t_i)\cos\theta_i)^2 + (\sum_{i=1}^{n} x(t_i)\sin\theta_i)^2] \quad (6)$$

We finally tabulate the considered T's and corresponding Z's for ultimate analysis. The values of T giving significant peaks for Z are considered to be the probable periods.

## 3. Results

| Data | Period (in months) |
|---|---|
| 1) SAGE data (1st January 1990 to 31st December 2000) | 1.3, 4.3, 5.5, 6.3, 7.9, 8.7, 15.9, 18.7, 23.9, 32.9 and 48.7. |
| 2) SAGE data (April 1998 to December 2001) | 1.5, 2.9, 4.5 and 10.1. |
| 3) GALLEX data (May 1991 to January 1997) | 1.7, 18.7 and 26.9. |
| 4) GNO data (May 1998 to December 2001) | 3.5, 5.5, 7.7 and 10.5. |
| 5) GALLEX-GNO combined data (May 1991 to December 2001) | 1.7, 18.5, 28.5 and 42.1. |
| 6) Average of the data from GNO and SAGE (May 1998 to December 2001) | 4.3, 6.9, 10.3 and 18.1. |

(See the figures 1, 2, 3, 4 for illustration for the data 1), 2), 3) and 4).)

## 4. Discussion

Earlier we obtained periodicities of solar neutrino flux data from SAGE and GALLEX-GNO detectors by Ferraz-Mello method of DCDFT and by Periodogram method [15]. For (1) the obtained period of 18.7 and 23.9 months by the present method are appreciably similar to the periods for (1) obtained by Ferraz-Mello method as 19.007 and 23.720 months and by Periodogram method as 18.614 and 23.851 months. For (2) the obtained periods of 1.5 and 2.9 months by the present analysis are similar to the period 1.501 months obtained by Ferraz-Mello and 1.504, 2.993 months obtained by Periodogram method. For (3) the obtained periods of 1.7 and 18.7 months are similar to the periods 1.647, 18.890 months by Ferraz-Mello method and 1.647, 18.635 months by Periodogram method. For (4) the obtained period of 3.5 months are similar to the period 3.399 months by Ferraz-Mello method and 3.433 months by Periodogram method. For (5) the obtained periods of 1.7 and 18.5 months are similar to the periods of 1.694, 19.002 months by Ferraz-Mello method and 1.694, 18.507 months by Periodogram method. The observed periods of 1.3 months for (1), 1.5 months for (2), 1.7 months for (3) and 1.7 months for (3) and 1.7 months for (5) fall within the region of 10-60 days periodicity estimated by Sturrock and Scargle [16].

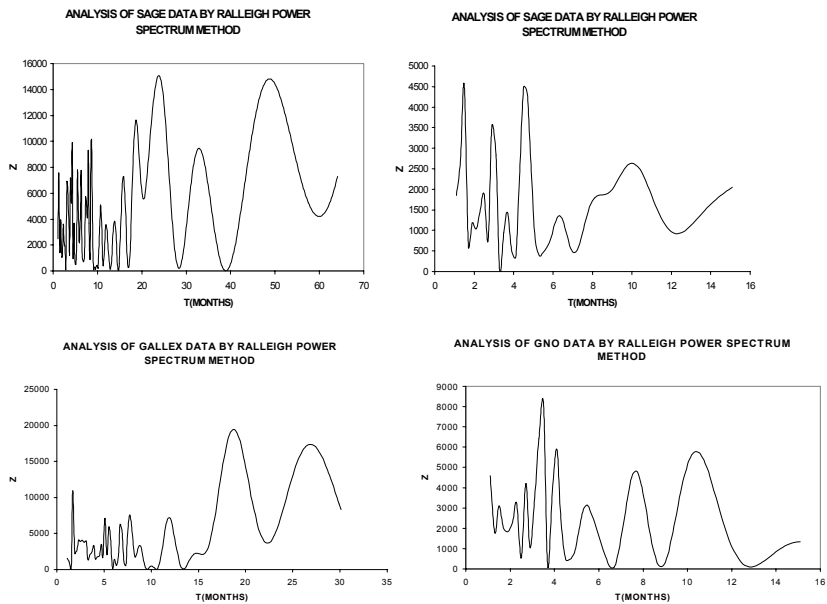

**Figure.1:** SAGE data (January 1990 to December 2000), **Figure.2:** SAGE data (April 1998 to December 2001), **Figure.3:** GALLEX data (May 1991 to January 1997), **Figure.4:** GNO data (May 1998 to December 2001).